\documentclass[epj]{svjour}
\usepackage{graphics}
\usepackage{amsmath}
\usepackage{graphicx,epsfig,psfrag}
\usepackage{amssymb}

\newcommand{\OLD}[1]{{\bf OLD RM}}

\renewcommand{\vec}[1]{{\mathbf #1}}

\newcommand{\w}{\omega}

\begin{document}
\title{Random Field effects in field-driven quantum critical points}
\author{Fabrizio Anfuso and Achim Rosch}
\institute{Institute for Theoretical Physics, University of Cologne, 50937
Cologne, Germany}
\date{Received: date / Revised version: date}
%
\abstract{  We investigate the role of disorder for field-driven
  quantum phase transitions of metallic antiferromagnets. For systems
  with sufficiently low symmetry, the combination of a uniform
  external field and non-magnetic impurities leads effectively to a
  random magnetic field which strongly modifies the behavior close to
  the critical point.  Using perturbative renormalization group, we
  investigate in which regime of the phase diagram the disorder
  affects critical properties. In heavy fermion systems where even
  weak disorder can lead to strong fluctuations of the local Kondo
  temperature, the random field effects are especially pronounced.  We
  study possible manifestation of random field effects in experiments
  and discuss in this light neutron scattering results for the field
  driven quantum phase transition in CeCu$_{5.8}$Au$_{0.2}$.
} 
\maketitle

\section{Introduction}
Disorder effects can strongly modify critical properties close to phase transitions. When investigating the role
of disorder, one usually distinguishes two cases: in random-field systems, the disorder couples {\em linearly}
to the order parameter, while in so-called random mass systems, the coupling is to the square of the order parameter.

Random fields effects are by far more dramatic compared to the
random-mass case. As has been shown in a seminal paper by Imry and Ma
\cite{Imry1}, weak random field even destroys completely long range
order for magnets with xy or Heisenberg symmetry as the energy costs
to form domain walls are smaller than the energy gain when the
magnetic structure adapts locally to the random field. For magnets
with Ising symmetry, long range order is stable in three dimensions
(3D) as long as the random fields are weak but the properties close to
the phase transition are strongly modified.

Random field criticality, and especially the Random Field Ising
Model (RFIM), has attracted considerable interests in the last decades.
Despite this broad activity \cite{reviews} including numerical, analytical
 and experimental investigations, a complete understanding of the RFIM is still lacking.
It is established by now that the results of the perturbative RG
calculation (the
so-called "dimensional reduction" \cite{Aharony1,Grinstein}) are incorrect
and a consistent theoretical treatment should necessarily rely on some
sort of non-perturbative approach. Steps in this direction has been made
recently with the help of the functional renormalization group
\cite{Wiese,Tarjus1} but
the issue of the determination of the correct critical exponents is far from
being settled. Moreover, the intrinsic
"glassiness" of the RFIM renders the problem hard to be tackled
numerically \cite{Middleton1} and, to our
knowledge, an unified view of the critical behavior is not
yet available. All these complications also arise at a
quantum-critical point in the presence of random fields
as has been recently discussed by Senthil \cite{Senthil}.

An important step towards the experimental realization of RFIM was made by
Fishman and Ahrony \cite{Fishman} who suggested to
study doped Ising antiferromagnets in uniform
magnetic fields. Remarkably, they showed that the net effect of the
random moments induced by the non-magnetic doping plus the applied
field  naturally leads to
 the same critical behavior as the RFIM. This setup has the unique advantage
 that one can easily tune the effective strength of the random field
 just by changing the size of
 the external uniform field.
These observations paved the way for many interesting experiments
\cite{Belanger1} and for
further theoretical work. For example, in a series of recent
neutron scattering measures \cite{Ye1,Ye2}, the lightly-doped iron compound ${\rm Fe_{1-x}Zn_xF_2}$
showed all the expected features of the RFIM in and out of
equilibrium. For this material, an accurate experimental determination
of  critical exponents was possible.

Recently, also the effects of random-mass disorder close to quantum
phase transitions in metallic antiferromagnets have been studied
theoretically by a number of authors \cite{vojta}. Close to the
critical point, disorder leads to the formation of magnetic domains.
In contrast to the random-mass case these domains are not pinned by
external fields and can have stronger fluctuations. This has been
predicted to lead to pronounced Griffiths singularities, smeared
transitions and glassy states \cite{vojta}.

In the present paper, we investigate the role of random fields on {\em
  quantum} phase transitions of metallic magnets ignoring the (much
smaller) role of random mass effects. As in the setup
proposed by Fishman and Ahrony \cite{Fishman} for classical systems,
we consider field driven transition in weakly disordered metallic
magnets with Ising symmetry.  In these systems, where long range order
\emph{can} exist, the role of the magnetic field is two-fold. On the
one hand, coupled to the (non-magnetic) impurities, it generates the
random field; on the other hand, it induces a quantum phase transition
at a certain critical strength. Field-driven quantum phenomena are
today a major topic of experimental investigation, both in the context
of magnetic insulators, where the field typically induces a
Bose-Einstein condensation of magnons, and in the one of magnetic
metals, where the field instead leads to the suppression of the long
range order \cite{Inga}.

Our analysis addresses the latter case. As an example, we have in mind
the heavy-fermion compound ${\rm CeCu_{6-x}Au_x}$, for which many
experimental data are already available in the literature
\cite{Lohnesyan1,Lohnesyan2}.  This material
 is a prototypical heavy-fermion system governed by the competition
of Kon\-do scree\-ning and RKKY interactions. For a finite concentration
of gold and below a certain critical field, antiferromagnetic Ising
order appears at a finite N\'eel temperature $T_N$. For a doping
$x>0.1$ the metal orders magnetically at low temperatures. For a
doping larger than $0.1$, the magnetic order can be suppressed by a
uniform magnetic field which allows a precise study of a field-driven
quantum phase transition. The doping by Au atoms naturally induces
disorder. For example, a doping $x=0.2$ reduces \cite{Lohnesyan1} the effective Kondo
temperature on average by approximately 50\%. This implies
that also the Kondo temperatures will strongly vary locally.
Therefore one expects in the presence of a uniform magnetic field
rather strong fluctuations in the local magnetization. These
static spatial fluctuations play the role of effective random fields and are expected to
modify dramatically the properties close to the quantum phase
transition. For example, magnetic domains will start to nucleate even
on the non-magnetic side of the phase diagram. As we will discuss,
recent elastic neutron scattering results from Stockert {\it et al.}
\cite{Stockert} appear to be consistent with this scenario.

In this paper, we will not try to describe the critical properties of
field driven quantum phase transitions directly at or very close to
the quantum critical point. In this regime one has to face all the
(unsolved) problems well known from the classical RFIM as has been
shown in Ref.~\cite{Senthil}. Instead we
will focus on the much more simple, but nevertheless experimentally
relevant question, in what regime the quantum-critical properties of
the clean system are modified by the random field. What are the
signatures of the onset of random-field physics? To answer this
question we use standard perturbative renormalization group (RG)
methods to determine the properties of the phase diagram and the
location of the crossover lines. The perturbative approach is
combined with phenomenological considerations based on the Imry-Ma
\cite{Imry1} argument.


In the following, we first review the derivation of the RG equations
\cite{Grinstein,hertz,Micnas,Millis} and we obtain
the general phase diagram with the different physical regimes and
crossover lines. Then, we focus on equilibrium and
out-of-equilibrium experimental
quantities and we list possible smoking guns for the onset of
random field physics.

\section{Model and perturbative RG equations}
The starting point is the usual Landau-Ginzburg-Wilson functional for the
order parameter \cite{hertz}
\begin{eqnarray}\label{S}
\emph{S}(m)&=&\frac{1}{2}\sum_{\lambda}\Bigl[\Bigl
(\delta+q^2+\frac{|\omega_n|}{\Gamma}\Bigr )m_{\lambda}m_{-\lambda}+h_{\mathbf{q}}m_{\lambda}\delta_{\omega_n,0}\Bigr]
\nonumber\\ &&+\frac{u_0}{\beta}\sum_{\lambda,\lambda',\lambda''}m_{\lambda}m_{\lambda'}m_{\lambda''}
m_{-\lambda-\lambda'-\lambda''}
\end{eqnarray}
where $m$ is the Ising order parameter,
$\lambda=(\omega_n,\mathbf{q})$, $\Gamma$ is a characteristic energy scale
(set to $\Gamma=1$ in the following)
 and $\delta$ is the critical
tuning parameter (proportional to $B-B_c$ for a transition driven by a uniform field $B$).
The momentum $\vec{q}=\vec{k}-\vec{Q}$ is measured with respect to the ordering vector $\vec{Q}$
of the antiferromagnet and the $|\omega_n|$ describes the
damping of spin fluctuations by a coupling to particle-hole pairs in a metal. Due to its presence, typical energies
scale as $q^2$ in the clean system and therefore the dynamical critical exponent is $z=2$. For an insulator (or
a metal with small Fermi surface, $2 k_F < Q$) the $|\omega_n|$ is replaced by a $\omega_n^2$ term such that $z=1$
in this case.

$h(x)$ is the static random
field that is
assumed to be Gaussian correlated,
\begin{equation}
\langle
h(\mathbf{q})\rangle=0\qquad{\rm and}\qquad
\langle
h(\mathbf{q})h(\mathbf{q}')\rangle=h^2\delta(\mathbf{q}+\mathbf{q}'),
\label{disorderdistribution}
\end{equation}
with tunable strength $h$ which is typically proportional
to the strength of non-magnetic disorder and, more importantly, to the external uniform field.

\begin{figure}
\includegraphics[width=.45\textwidth,clip]{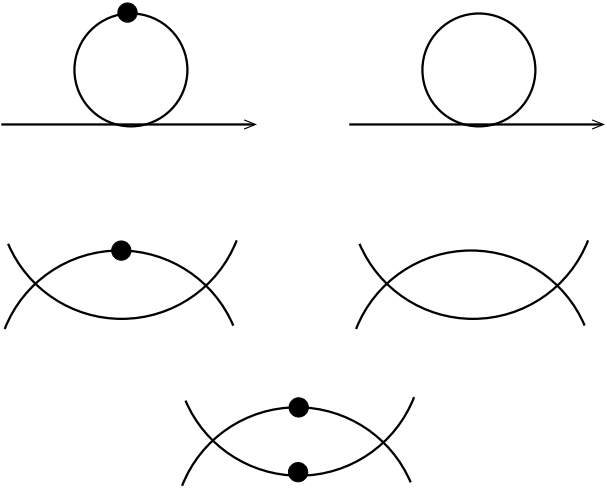}
\caption{The one-loop diagrams in the presence of disorder. The black
  dot indicates the random field contribution.}
\label{Feynman}
\end{figure}

With the help of the standard replica trick, we can average over the
disorder replicating the action
\begin{eqnarray}
\emph{S}(m^{\alpha})&=&\frac{1}{2}\sum_{\lambda,\alpha,\alpha'}\Bigl[\Bigl(\delta+q^2+\frac{|\omega_n|}{\Gamma}\Bigl)\delta_{\alpha,\alpha'}-\beta
h^2\delta_{\omega_n,0}\Bigr]\times\nonumber \\&\times& m_{\lambda}^{\alpha}m_{-\lambda}^{\alpha'}
+\frac{u_0}{\beta}\sum_{\lambda,\lambda',\lambda''}m_{\lambda}^{\alpha}m_{\lambda'}^{\alpha}m_{\lambda''}^{\alpha}
m_{-\lambda-\lambda'-\lambda''}^{\alpha}\qquad\label{replicatedaction}
\end{eqnarray}
where $\alpha=1,2..,n$ are the replica-indices. We have
formally eliminated the random field and, in absence of the quartic interaction,
the free propagator is given by
\begin{eqnarray}
\langle
m^{\alpha}_{\lambda}m^{\alpha'}_{\lambda'}\rangle&=&\delta_{\lambda,-\lambda'}\Bigl(\delta+q^2+\frac{|\omega_n|}{\Gamma}\Bigl)^{-1}\Bigl[\delta_{\alpha,\alpha'}
+\nonumber\\&+&\beta
h^2\delta_{\omega_n,0}\Bigl(\delta+q^2\Bigr)^{-1}\Bigr].\label{propagator}
\end{eqnarray}
Observe that already within the Gaussian approximation the propagator at $\w_n=0$ is highly singular in the presence
of the random field. Computing the
diagrams in Fig.~\ref{Feynman} for the replicated action in
Eq.~(\ref{replicatedaction}), we obtain the following closed set of one-loop RG equations
(similar to the one obtain by Micnas and Chao \cite{Micnas})
\begin{eqnarray}
\frac{d\delta}{d\log b}&=&2\delta+6u\emph{f}_1(T)+6vf_2 \label{RG1}\\
\frac{dv}{d\log b}&=&(6-d)v-36uv\emph{f}_3(T)-72v^2\emph{f}_4\label{RG2}\\
\frac{du}{d\log b}&=&[4-(d+z)]u-36u^2\emph{f}_3(T)-72uv\emph{f}_4\label{RG3}\\
\frac{dT}{d\log b}&=&zT(b)\label{RG4}
\end{eqnarray}
where we defined $v=uh^2$ as a renormalized disorder parameter and we
introduced the functions $\emph{f}_n$ ($n=1,2,3,4$) defined in the appendix. Note that while $f_1$ and $f_3$
depend both on the renormalized $\delta(b)$ and $T(b)$, the functions $f_2$ and $f_4$ depend
on $\delta(b)$ only as the disorder is static.
The upper critical dimension $d=6$ is set by the renormalized disorder
$v$. As the classical RG equations \cite{Grinstein}, Eqns.~(\ref{RG1}-\ref{RG4}) have to leading order in $\epsilon=6-d$
 a stable fixed point, $v^*=\epsilon/72$, $u^*=0$. The critical exponents appear to be
  given by the ones of the equivalent clean system in dimensions  $d^*=d-2$ (dimensional reduction, see above).

\section{Phase diagram}
As already pointed out in the introduction, the pertubative fixed point and the
critical exponents are incorrect but they are not our main focus here. We
consider instead small bare couplings, $h^2\ll 1$ and $u\ll
1$, studying the RG equations in the vicinity of the quantum critical point of the clean system, described
by an
unstable Gaussian fixed point. Following the flow of the coupling
constants, we identify the different
crossover lines by investigating for which parameters $\delta$, $v$ or $u$ become of order $1$.
Linearizing Eqns.~(\ref{RG1}-\ref{RG4}) around $u=v=\delta=0$, we obtain
\begin{eqnarray}
\frac{d\delta}{d\log b}&=&2\delta+6 u \emph{f}_1(T)+6 v \emph{f}_2\\
\frac{du}{d\log b}&=&[4-(d+z)]u\\
\frac{dv}{d\log b}&=&(6-d)v\\
\frac{dT}{d\log b}&=&zT,
\end{eqnarray}
where $f_2$ is now just a number,
and we can write the formal solution
\begin{eqnarray}
\delta(b)&=&b^2\Bigl[\delta_0+6u_0\int_0^{\log b}dx
e^{[2-(d+z)]x}\emph{f}_1(Te^{zx})\nonumber \\
&&+6v_0\emph{f}_2\int_0^{\log
  b}e^{(4-d)x}\Bigr]\label{RG1L}\\
u(b)&=&u_0b^{4-(d+z)}\label{RG2L}\\
v(b)&=&v_0b^{6-d}\label{RG3L}\\
T(b)&=&T_0b^{z}\label{RG4L}.
\end{eqnarray}
Here $u_0, v_0$ are the bare values of the couplings. Correspondingly,
$T_0$ is the bare, i.e. the physical temperature.
The features of the function $\emph{f}_1(T)$ have been already
discussed in \cite{Millis} (see appendix). For
our purposes, we only need to know that
\begin{eqnarray}
\emph{f}_1(T)&\approx&\emph{f}_1(0)\qquad T\ll 1\label{cond1}\\
\emph{f}_1(T)&\approx&CT\qquad T\gg 1\label{cond2}
\end{eqnarray}
\begin{figure*}
\begin{center}
\includegraphics[width=0.85 \textwidth]{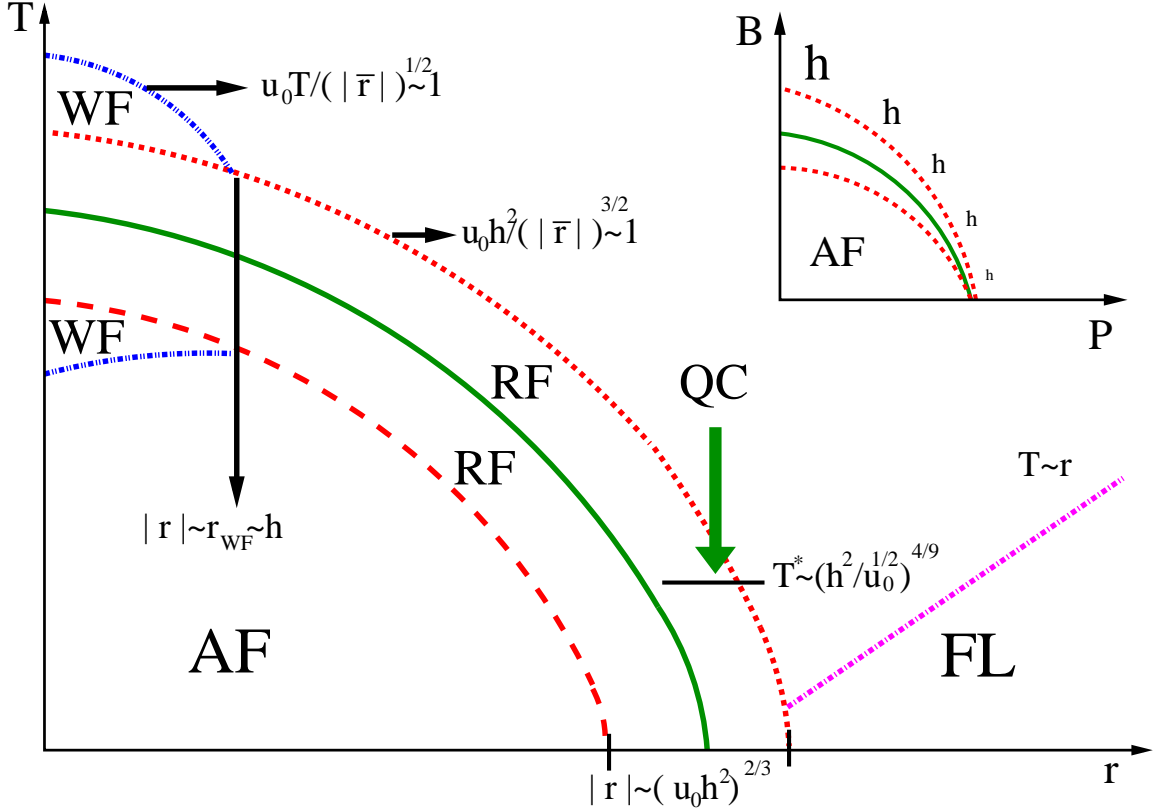}
\end{center}
\caption{Main figure: Phase diagram for a quantum critical
  antiferromagnetic metal  with Ising symmetry  in $d=3$
as a function of temperature and control parameter $r \propto B-B_c$ in
  the presence of random field disorder $h$.
Close to the phase transition line, where long-range order vanishes
(solid green line), the effects of disorder become dominant (region enclosed
by the two dotted red lines parallel to the solid green line).
The crossover from the  quantum critical to the Fermi liquid regime
of the underlying clean system upon lowering $T$ occurs at $T\sim r$
(dotted magenta line). The Wilson Fisher
regime of the classical phase transition in the clean limit is only
present for $r<-h$ (dotted blue  lines at the left part of the
diagram). Inset: Phase diagram of a system
(e.g. $\rm CeCu_{5.8}Au_{0.2}$)
where magnetism can be suppressed either by pressure or an applied
magnetic field. As the strength of the effective disorder is
  proportional to the applied uniform field, $h \propto B$, one can
 tune the effective strength of random-field disorder.}
 \label{phasediagram}
\end{figure*}
We first investigate the low-temperature ``Fermi liquid'' (FL) regime on the disordered
side of the phase diagram, see Fig.~\ref{phasediagram}. Here, $\delta(b)$ reaches the cutoff first
in a regime where $T(b)\ll 1$ and $v(b) \ll 1$.
Using Eq.~(\ref{cond1}) and expanding Eq.~(\ref{RG1L}), we obtain for
$d+z>4$ (we are mainly interested in the metallic, three-dimensional case, $d=3, z=2$)
\begin{eqnarray}
1&=&b^2[\delta_0+\frac{6u_0\emph{f}_1(0)}{d+z-2}+6v_0\emph{f}_2\frac{(b^{(4-d)}-1)}{4-d}]\nonumber\\
&&\approx b^2[\delta_0+\frac{6u_0\emph{f}_1(0)}{d+z-2}]=b^2r\\
T&=&T_0b^z=T_0/r^{z/2}\ll 1 \label{tR}\\
v&=&v_0r^{(d-6)/2}\ll 1 \label{vR}
\end{eqnarray}
where $r$ describes the distance from the quantum-critical point of
the underlying {\em clean} system,
\begin{equation}
r=\delta_0-\delta_0^c=\delta_0+\frac{6u_0\emph{f}_1(0)}{d+z-2} \label{criticalT0}.
\end{equation}
The
first inequality (\ref{tR}) describes the crossover to the quantum
critical regime, see Fig.~\ref{phasediagram}, while
the second  (\ref{vR}) the onset of a regime, where the random field
dominates the critical behavior
(see discussion below).

A similar condition for the onset of random-field criticality can be
obtained in the ordered phase. Here we use the phenomenological
Imry-Ma argument \cite{Imry1}.
The random field favors the proliferation of magnetic domain walls
where each domain (of radius $L$) adjusts itself to the underlying random-field landscape.
Far enough from the transition, the energy cost of a
domain wall \cite{Chaikin} is proportional to
$\frac{|r|^{3/2}}{u}L^{d-1}$ for $L$ large compared to the correlation
length $\xi$. The energy gain from the random field
scales with $h m L^{d/2}$ where the size of the order parameter is
given by $m=(|r|/u_0)^{1/2}$.
In $d>2$, the size of a random-field induced domain is therefore
limited by
\begin{equation}
h\Bigl(\frac{|r|}{u}\Bigr)^{1/2}L^{d/2}>\frac{|r|^{3/2}}{u} L^{d-1}.
\end{equation}
This inequality has an solution with $L>\xi\approx r^{-1/2}$ only for
\begin{equation}
|r| \lesssim  (uh^2)^{\frac{2}{6-d}}. \label{imrymaginzburg}
\end{equation}
which is equivalent to the condition (\ref{vR}) obtained from
perturbative RG for the disordered side of the phase
diagram. Therefore if $|r|$ is sufficiently small, magnetic domains
caused by the random field will proliferate while deep in the ordered
phase only exponentially rare random-field configurations induce (small) domains.

In $d=2$, the long range magnetic order is always destroyed and
fragments into domains. In this case, (\ref{imrymaginzburg}) describes the condition for a crossover
from exponentially large domains for large negative $r$ to a disorder dominated regime for smaller $r$ \cite{binder}.

The quantum critical (QC) regime is obtained when approaching the
quantum phase transition by lowering the temperature. In this regime,
the  rescaled $T(b)=T_0 b^z$ reaches the cutoff, $T(b=b^*)=1$ where
the RG flow changes its nature (but is not stopped).
 Therefore it
 is useful to split the RG flow in two steps,  rewriting
 Eq.~(\ref{RG1L}) as follows
\begin{eqnarray}
\delta(b)&=&b^2\Bigl[\delta_0+6u_0\int_0^{\log b^*(T)}dx
e^{[2-(d+z)]x}\emph{f}_1(Te^{zx})\nonumber \\
&&+6u_0\int_{\log b^*(T)}^{\log b}dx
e^{[2-(d+z)]x}\emph{f}_1(Te^{zx})\nonumber\\ &&+6v_0\emph{f}_2\int_0^{\log
  b}e^{(4-d)x}\Bigr]\label{RG1Lbis}
\end{eqnarray}
where we defined $b^*(T_0)=T_0^{-1/z}$. Using Eqns.~(\ref{cond1}) and (\ref{cond2}), we can perform the integrals in Eq.~(\ref{RG1Lbis}),
to obtain
\begin{eqnarray}
\delta(b)&=&b^2\Bigl[r+6u_0\bigl(\frac{\emph{f}_1(0)}{2-(d+z)}+\frac{C}{d-2}\bigr)T^{\frac{d+z-2}{z}}\nonumber\\
&&+\frac{6v_0\emph{f}_2}{4-d}b^{4-d}\Bigr].\label{cri}
\end{eqnarray}
The combination
\begin{equation}
\bar{r}=r+6u_0\bigl(\frac{\emph{f}_1(0)}{2-(d+z)}+\frac{C}{d-2}\bigr)T^{\frac{d+z-2}{z}},
\end{equation}
can be identified with the distance from the finite-temperature phase
transition line \cite{Millis}.

To find the crossover lines to the random-field dominated regime, we
have to investigate under which condition, the term proportional to
$v_0$ dominates in (\ref{cri}) of the clean system. For $v_0=0$,
$\delta(b)$ reaches $1$ for $\bar b=\bar{r}^{-1/2}$. As the RG flow stops
for $\delta(b)>1$, we can obtain the crossover line from the condition
$\bar{r}\approx \frac{6v_0\emph{f}_2}{4-d}{\bar b}^{4-d}$.
Random field physics therefore dominates for
\begin{equation}
\bar r \lesssim (u_0 h^2)^{\frac{2}{6-d}}. \label{ginzburgdisorder}
\end{equation}
This is also the condition that $v(\bar b)$ becomes of order $1$ (note
that $\bar b$ and not $b^*$ enters here).
 This last equation is \emph{de
  facto} a Ginzburg criterion for the on-set of the random field effects.
Notice that Eq.~(\ref{ginzburgdisorder}) is the extension to finite
temperature of Eq.~(\ref{imrymaginzburg}) that followed directly from
the Imry-Ma argument.

For $r=0$, i.e. in the quantum-critical region of the
underlying clean system, the effects of random fields are weak for
$T_0>T^*$ with
\begin{eqnarray}
T^{*}&\approx& u^{\frac{z(d-4)}{(6-d)(d+z-2)}}h^{\frac{4z}{(6-d)(d+z-2)}}\nonumber\\
&\approx& \Bigl(\frac{h^2}{u^{1/2}}\Bigr)^{4/9} \qquad {\rm for } \quad  d=3,z=2.
\end{eqnarray} according to Eq.~(\ref{ginzburgdisorder}).

The calculations given above are obviously not valid very close to the
{\em classical} phase transition of the clean system where also $u$
is relevant. Very close to the QCP this Wilson-Fisher critical regime
is never reached because the random field becomes relevant first.
Comparing the Ginzburg criterium for the clean system
\begin{equation}
\frac{uoT_0}{\bar{r}^{(4-d)/2}}\approx 1.\label{WFcrossover}
\end{equation}
with Eq.~(\ref{ginzburgdisorder}), we find that the system will enter
the Wilson-Fisher regime of the clean system {\em before} the random
fields becomes relevant only for $|r| > r_{WF}$ (see
Fig.~\ref{phasediagram}) with
\begin{eqnarray}
r_{WF}&\approx& h^{\frac{8-2 d}{6-d} \, \frac{d}{z}}
u_0^{\frac{6 z -    d z -2 d}{(6-d) z}} \nonumber \\
 &=& h \qquad {\rm for } \quad  d=3,z=2
\end{eqnarray}

Combining the results for the various crossover lines, we obtain the
phase diagram shown in Fig.~\ref{phasediagram}.


\section{Experimental probes}

As pointed out in the introduction, field driven quantum phase
transitions offer a high degree of tunability. An external magnetic
field changes not only the distance
to the quantum-critical point but also the effective strength of
the random-field disorder, $h \propto B$. Ideally, one can use a
combination of
doping and magnetic field or, even better, of pressure and magnetic field to
tune the random field strength and the distance from the QCP
independently. For example, considering a system like
CeCu$_{5.8}$Au$_{0.2}$, the quantum critical point can be reached
either by applying moderate magnetic fields or a moderate
pressure. By crossing the phase transition line in the $B,p$ plane for
different values of the magnetic field (see inset of Fig.~\ref{phasediagram}), one can systematically study
how random fields modify the physics.

\subsection{Elastic neutron scattering and comparison to CeCu$_{5.8}$Au$_{0.2}$}\label{neutronscattering}
Elastic neutron scattering probes the static magnetization of the
sample as a function of momentum. In a random field system there
is always a finite, spatially-fluctuating static
magnetization even far away from the transition. In contrast, in a
clean system (or in the presence of only random-mass disorder), a finite-momentum static
magnetization appears only in the ordered phase \cite{footnote}.
\begin{figure}
\includegraphics[width=.50\textwidth,clip]{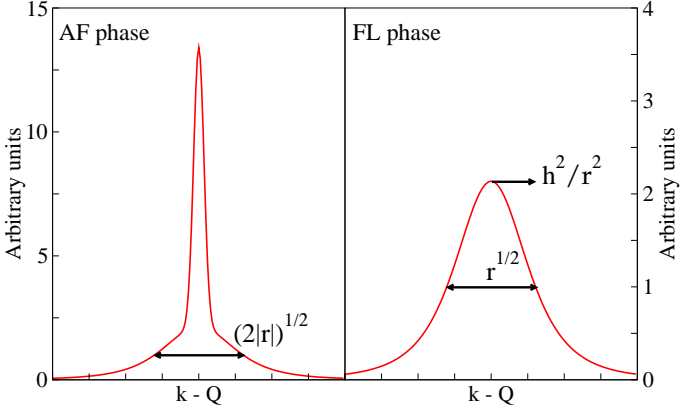}
\caption{Sketch of the elastic neutron scattering signal in the
  ordered and disordered phases. Deep in the ordered phase, one expects
  a resolution limited peak of Gaussian shape
at the ordering wave vector. Random fields induce a double-Lorenzian
tail of width $\sqrt{2|r|}$, see Eq. (\ref{sigmael}).
Upon approaching the random-field regime,
the weight of the tail becomes of the order of the weight of the
central peak. Also in the Fermi liquid regime, the random field gives
a contribution to the neutron scattering signal described
by Eq.~(\ref{sigmael2}).
\label{neutronSchematic}}
\end{figure}
Outside of the random-field regime, i.e for $|r|>(u_0 h^2)^{2/3}$
(assuming $d=3$ in this chapter)
one can easily estimate the effects of the random field. Even deep in
the antiferromagnetic phase, the random fields will induce an extra
contribution to the magnetization, $\delta m_q \approx \chi_q h_q$,
where $\chi_q \approx 1/(-2r +q^2)$ is the susceptibility of the
clean system in the ordered phase ($r<0$) and $q$
measures the distance from the ordering wave vector.
 Averaging over
disorder, we obtain a corresponding contribution to elastic neutron
scattering.
The total elastic cross section
\begin{eqnarray}
\sigma_{el}(q) \approx m_0^2 \,\, \delta^3(q)+
\frac{h^2}{(|2 r|+q^2)^2}
\label{sigmael}
\end{eqnarray}
therefore obtains a contribution both from the long-range order and
from the random fields. There is also an extra contribution (not shown
above) from exponentially rare field-induced domains and the
corresponding domain walls. These domains proliferate upon approaching
the random-field regime. A schematic picture of the expected neutron
signal (including resolution effects) is shown in Fig.~\ref{neutronSchematic}.

Similarly, approaching the transition from the disordered side one
obtains
\begin{eqnarray}
\sigma_{el}(q) \approx
\frac{h^2}{( r+q^2)^2}
\label{sigmael2}
\end{eqnarray}
As above, there are some rare regions where the random fields are
strong and for which (\ref{sigmael2}) cannot be used but these
give only sub-leading contributions for
 $r>(u_0 h^2)^{2/3}$.

For an experiment with finite resolution, it is not possible to separate
the two contributions in (\ref{sigmael}) very close to the
transition. In the random-field regime, the weight of the order
parameter is always smaller than the weight of the disorder contribution.
As a consequence, the phase transition appears to be
'smeared out'.

Qualitatively, such a behavior has been observed in recent neutron
scattering experiments on the field-driven QPT in
CeCu$_{5.8}$Au$_{0.2}$ by Stockert {\it et al.}
\cite{Stockert}. Upon approaching the transition, the elastic neutron
scattering signal broadens and the weight of the signal is reduced.
However, there are no sharp features and the integrated intensity
remains finite well beyond the critical field. According to
Eq.~(\ref{sigmael2}), the width of the signal should grow
as $\sqrt{r} \sim \sqrt{B-B_c}$ while the total integrated intensity
vanishes slowly with $h^2/\sqrt{r}$ (note that in Ref.~\cite{Stockert}
the intensity was obtained by integrating over a rocking scan, i.e. a
line in momentum space, leading to a more rapid decay $\sim
h^2/r^{3/2}$). While the experimental results appear to be consistent
with the predicted scenario, a fully quantitative comparison is not
possible due to the limited statistics and momentum
resolution of the experiment. For a quantitative comparison it would be useful to
consider samples with different doping and different critical fields.
Ideally, one would like to use both pressure and field to explore the
phase diagram sketched in the inset of Fig.~\ref{phasediagram} using a
single sample with fixed bare disorder strength. In such a case, the
random field $h$ will be roughly proportional to the critical field $B_c$.

\subsection{NMR}\label{NMR}
As a local probe,
nuclear magnetic resonance (NMR) is a natural tool to investigate systems with
magnetic textures and inhomogeneities. It is possible to measure
directly the distribution of the local magnetization
$M(\vec{r})=M_0+\delta M$.
Assuming a Gaussian distribution of the random fields,
$p[h(r)]\sim \exp\!\!\left(- \int h(r)^2/(2 h^2)\right)$ as in
Eq.~(\ref{disorderdistribution}), we can derive the corresponding
distribution of magnetic moments outside of the random field regime
using again $\delta m_q \approx \chi_{\mathbf{q}} h_q$ to obtain
\begin{eqnarray}
P(\delta M)&\approx& \exp\!\!\left(
-(\delta M)^2/(2 \sigma^2)\right)\label{stagmoments}\\
\sigma^2&=& h^2 \int\chi_{\mathbf{q}}^2\sim
\frac{h^2}{r^{(4-d)/2}}
\end{eqnarray}
For a field-driven quantum phase transition in $d=3$, we therefore expect a
width $\sigma$ proportional to $B/(B-B_c)^{1/4}$.

\subsection{Non-equilibrium effects}

Probably the most direct way to detect random-field physics, is the
observation of hysteretic  behavior in the random-field regime which shares
similarity to the properties of (spin-) glasses. This
is well known both theoretically and experimentally for classical random-field
systems. Upon approaching the classical critical point, the relevant relaxation
rates grow exponentially \cite{Villain,Fisher,Nattermann} with the correlation length,
\begin{equation}\label{rate}
\tau \sim \exp[c \, \xi^{\theta}/T],
\end{equation}
where $\theta$ parameterizes the typical free energy of the system at
the scale $\xi$, $F_\xi\sim \xi^\theta$. For conventional critical
points one has $F_\xi\sim T \sim const.$. A
divergence of $F_\xi$ leads to a violation of hyper-scaling,
$(d-\theta)\nu=2-\alpha$ ($\nu$ and $\alpha$ are the
correlation length and specific heat exponents). Therefore $\theta$
describes to what extent there is an effective dimensional reduction,
$d \to d-\theta$. In Ref.~\cite{Middleton1}, $\theta\approx 1.49 \pm
0.03$ was estimated numerically for $d=3$.

\begin{figure}
\includegraphics[width=.36\textwidth,clip]{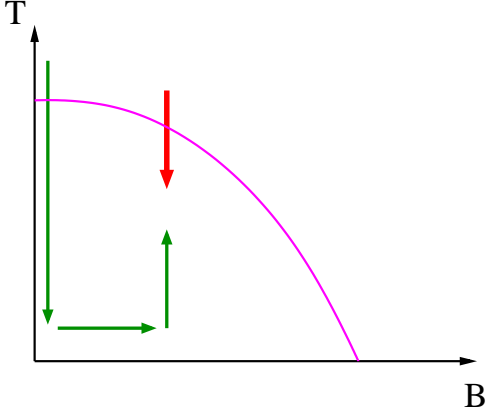}
\caption{The two non-equivalent cooling procedures. The zero-field
  cooling path is indicated with the thin green arrows while the
field cooling scheme uses a path along the
 thick red arrow.}
\label{magnetization}
\end{figure}

Experimentally, the long relaxation rates lead to strong
non-equilibrium effects.  Neutron scattering experiments
\cite{Ye2,Belanger1} and
magnetization measurements \cite{Kleemann} on diluted antiferromagnets
in uniform magnetic fields show hysteretic behavior. It is, for
example, useful to compare field-cooled with zero-field cooled
samples. Zero field cooling (green
arrows in Fig.~\ref{magnetization}) leads to a well-ordered
antiferromagnetic state. Instead, field cooling causes a fragmentation
of the antiferromagnet into non-equilibrated domains. The size of the
domains depend on the strength of the random field \cite{Villain} and
logarithmically on the cooling rate, see
Eq.~(\ref{rate}).
For example, experiments in field-cooled and
zero-field-cooled $\rm Fe_{0.85}Zn_{0.15}F_2$, give distinct critical
behavior \cite{Ye2,Belanger1}. Moreover, sweeping fast in
temperature back and forth through the transition, can cause the onset
of hysteresis phenomena in the intensity of the Bragg peak.

From energetic and renormalization group arguments, it is believed
\cite{reviews,Senthil} that the static properties of the $T=0$ and
finite-temperature random-field transitions are described by the same
fixed point. The dynamics and therefore the non-equilibrium properties
can, however, differ.
For $T \to 0$, the dynamics is not any more governed by thermal
fluctuations and thermal activation but instead by quantum tunneling.  Nevertheless, the
existence of extensive energy barriers, $E_\xi \sim \xi^\theta$, will
lead again \cite{Senthil} to exponentially long relaxation times
 \begin{equation}\label{rate1}
\tau \sim \exp[c' \, \xi^{\Psi}],
\end{equation}
 with $\Psi \neq \theta$ for insulators.
In a metal, one has furthermore to take into account, that
quantum-tunneling can be suppressed \cite{Millis1,vojta} by the coupling of
the order parameter to the electrons. This will lead to an additional
suppression of the tunneling rate proportional to $T^{c'' \xi^\gamma}$ 
 \begin{equation}\label{rate1}
\tau \sim T^{-c'' \xi^\gamma} \exp[c' \, \xi^{\Psi}],
\end{equation}
with a new (and unknown) exponent $\gamma$
such that $\tau(T=0)=\infty$ for sufficiently large domains.

Note that long relaxation times, smeared transitions and glassy effects can also arise in
the absence of magnetic fields, i.e. in systems with random-mass
disorder \cite{vojta}. In general, these effects are expected to be
much weaker (the bare scaling dimension of the random-mass disorder is 
$4-d$ rather than $6-d$). Furthermore, the (bare) strength of the
random-mass disorder does not  depend strongly on the
strength of the external magentic field which allows to separate the
two effects.

\section{Conclusions}

Field-driven quantum phase transitions of antiferromagnets
offer a unique possibility to study the interplay of disorder and
quantum criticality. Here one has to distinguish two cases: when the
staggered magnetization is perpendicular to the applied field (the
typical situation for a system with Heisenberg symmetry) and the
opposite situation (for Ising symmetry). In the first case,
disorder effects are usually weak, the disorder couples only to the
square of the order parameter.  This paper focuses on the second case,
where effectively a random field  is generated \cite{Aharony1} which
couples linearly to the order
parameter. In easy plane antiferromagnets, one can have
both situations depending on the direction of the magnetic field.

As magnetic domains can nucleate and pin at random field
configurations, the critical behavior of the system is radically
changed by the presence of disorder. By comparing quantum critical
points of the same material at different pressure, different doping or
different field direction, one can directly observe how random-fields
of different strengths affect quantum criticality.

With the help of perturbative renormalization group we have derived the
generic features of the phase diagram for metallic Ising antiferromagnets in
weak random fields. While this method is not able to investigate the
regime, where random-field physics dominates, one can nevertheless
extract the relevant crossover scales. This is especially of interest
for the study of systems where the effective strength of random-field
disorder can be tuned as described above and in the inset of
Fig.~\ref{phasediagram}.

One effect of random fields is that the phase transition appears to be
``smeared''. We argue that this has been observed in recent elastic
neutron scattering experiments in CeCu$_{5.8}$Au$_{0.2}$ where the
elastic neutron scattering peak broadens and slowly diminishes in
weight when the magnetic field is increased beyond the critical field.

The most direct way to detect random-field physics is the observation
of hysteretic behavior as a function of temperature and, especially, field. In
metals, these non-equilibrium effects are enhanced compared to their
classical counterparts as quantum tunneling is inhibited by the
coupling to the particle-hole pairs of the metal and thermal activation
is suppressed by the low temperatures.

We hope that the high tunability of the random-field
physics will motivate further theoretical and especially experimental
studies of this very interesting problem.

\begin{acknowledgement}
We thank T. Nattermann, N. Shah, O. Stockert, M. Vojta and K. Wiese for useful discussions
and the research group 960 and the
SFB/TR 12 of the DFG for financial support.
\end{acknowledgement}

\begin{appendix}
\section*{Appendix A: $f_n$ functions}\label{f-functions}
Referring to the case of an
itinerant antiferromagnet, in accordance with \cite{Millis}, the $f_1$ function is defined as
\begin{eqnarray}
f_1&=&\frac{\Lambda^{d+z+2}\Omega^d}{\pi}\int_0^{\Gamma\Lambda^{z-2}} d\omega\coth\frac{\omega}{2T}\frac{\omega}{\Lambda^{2z}(\delta+\Lambda^2)^2+\Lambda^4\omega^2}
\nonumber\\
&+&\frac{2\Gamma^2}{\pi}\int_0^{\Gamma}\frac{d^dq}{(2\pi)^d}\coth\frac{\Gamma}{2T}\frac{q^{2+z}}{q^4\Gamma^2+q^{2z}(q^2+\delta)^2}
\end{eqnarray}
where $\Lambda$ is the momentum cut-off, $\Gamma$ is the frequency
cut-off, $\Omega_d$ is the volume of the surface of unitary radius in
$d$ dimensions and
$z=2$ is the dynamical exponent. 
The other functions are instead given by \cite{Micnas}:
\begin{eqnarray}
f_2&=&\frac{\Omega^d \Lambda^d}{(2 \pi)^d} \frac{1}{(\delta+\Lambda^2)^2}\\
f_3&=&-\frac{\partial f_1}{\partial\delta}\\
f_4&=&\frac{\Omega^d \Lambda^d}{(2 \pi)^d} \frac{1}{(\delta+\Lambda^2)^4}.
\end{eqnarray}

\end{appendix}


\begin{thebibliography}{235}
\bibitem{Imry1} Y.~Imry and S.K.~Ma, Phys.~Rev.~Lett. {\bf 35}, 1399 (1975).
\bibitem{reviews}T. Nattermann in {\em Spin Glasses and Random
    Fields}, edited by A. P. Young (World Scientific, Singapure,
  1997). T. Nattermann and J. Villain, {\em  Random-field Ising
    systems - A survey on current theoretical views}, Phase
  Transition {\bf 11}, 5 (1988).
\bibitem{Aharony1} A.~Aharony, Y.~Imry and S.K.~Ma,
  Phys.~Rev.~Lett. {\bf 37}, 1364 (1976).
\bibitem{Grinstein} G.~Grinstein, Phys.~Rev.~Lett. {\bf 37}, 944
  (1976).
\bibitem{Wiese} K.J.~Wiese and P.~Le Doussal, Markov Processes
  Rel. Fields {\bf 13}, 777 (2007).
\bibitem{Tarjus1} G.~Tarjus and M.~Tissier, Phys. Rev. B {\bf 78},
  024203 (2008),  M.~Tissier and G.~Tarjus, Phys. Rev. B {\bf 78}, 024204 (2008).
\bibitem{Middleton1} A.A.~Middleton, D.~Fisher, Phys.~Rev.~B {\bf 65},
  134411 (2002).
\bibitem{Senthil} T. Senthil, Phys. Rev. B {\bf 57}, 8375 (1998).
\bibitem{Fishman} S.~Fishman and A.~Ahrony, J.~Phys.~C {\bf 12}, 729 (1979).
\bibitem{Belanger1} D.P.~Belanger, Braz.~J.~Phys. {\bf 30}, 682 (2000).
\bibitem{Ye1} F.~Ye {\it et al.}, Phys.~Rev.~Lett. {\bf 89}, 157202
  (2002).
\bibitem{Ye2} F.~Ye {\it et al.}, Phys.~Rev.~B {\bf 74}, 144431 (2006).
\bibitem{vojta}
T. Vojta, Phys. Rev. Lett. 90, 107202 (2003); V. Dobrosavljevic and E. Miranda
Phys. Rev. Lett. {\bf 94}, 187203 (2005); 
T. Vojta and J. Schmalian, Phys. Rev. B {\bf 72}, 045438 (2005); 
G. Schehr and H. Rieger, Phys. Rev. Lett. 96, 227201
(2006); J. A. Hoyos and T. Vojta, Phys. Rev. Lett. {\bf 100}, 240601 (2008).
\bibitem{Inga} I.~Fischer and A.~Rosch, Phys.~Rev.~B {\bf 71},
  184429 (2005).
\bibitem{Lohnesyan1} H.~von L\"ohneysen, A.~Rosch, M.~Vojta and
  P.~W\"olfle, Rev.~Mod.~Phys. {\bf 79}, 1015 (2007).
\bibitem{Lohnesyan2} H.~von L\"ohneysen, J.~Phys. Condens. Matter {\bf 8}, 9689 (1996).
\bibitem{Stockert} O.~Stockert, M.~Enderle and H.~v.~L\"ohneysen,
  Phys.~Rev.~Lett. {\bf 99}, 237203 (2007).
\bibitem{hertz} J. A. Hertz, Phys. Rev. B {\bf 14}, 1165 (1976).
\bibitem{Micnas} R.~Micnas and K.A.~Chao, Phys.~Rev.~B {\bf 30}, 6785
  (1984).
\bibitem{Millis} A.J.~Millis, Phys.~Rev.~B {\bf 48}, 7183 (1993).
\bibitem{binder} K. Binder, Z. Phys. {\bf B50}, 343 (1983).
\bibitem{Chaikin} P.M.~Chaikin and T.C.~Lubensky, \emph{Principles of
    condensed matter physics}, Cambridge (2000).
\bibitem{footnote}  More precisely, in an actual
experiment, the energy resolution is finite and can in principle also
pick up small contributions from slowly fluctuating domains very close
to the transition.
\bibitem{Villain}J.~Villain, Phys.~Rev.~Lett. {\bf 52}, 1543 (1984).
\bibitem{Fisher} J. Villain, J. Phys. (Paris) {\bf 46}, 1843 (1985).
\bibitem{Nattermann} T. Nattermann,
Phys. Stat. Sol.~B {\bf 129}, 153 (1985);
D.~Fisher, Phys.~Rev.~Lett. {\bf 56}, 416 (1986); T. Nattermann, Phys. Rev. Lett. {\bf 61}, 223 (1988).
\bibitem{Kleemann} W.~Kleemann, Int.~J.~Mod.~Phys.~B, {\bf 7}, 2469 (1993).
\bibitem{Millis1} A. J. Millis, D. K. Morr, and J. Schmalian,
  Phys. Rev. Lett. 87, 167202 (2001).


\end{thebibliography}
\end{document}